\documentclass[a4paper,11pt]{article}
\usepackage[utf8]{inputenc}
\usepackage[T1]{fontenc}
\usepackage{authblk}
\usepackage{todonotes}
\usepackage{natbib}
\usepackage{hyperref}
\usepackage{float}
\usepackage{longtable}
\usepackage{setspace}
\usepackage{lineno}
\usepackage{amsmath}
\usepackage{pdflscape}

\usepackage{tabularx,ragged2e,booktabs,caption,siunitx}

\usepackage[margin=1in]{geometry}

\providecommand{\keywords}[1]
{
  \small
  \centerline{\textbf{\textsc{Keywords}}} #1
}

\title{Machine learning event detection workflows in practice: A case study from the 2019 Durrës aftershock sequence.}

\author[1,*]{Jack Woollam}
\author[1,]{Vincent van der Heiden}
\author[1]{Andreas Rietbrock}
\author[2,]{Bernd Schurr}
\author[2,4]{Frederik Tilmann}
\author[3]{Edmond Dushi}

\affil[1]{\footnotesize{Geophysical Institute (GPI), Karlsruhe Institute of Technology, Karlsruhe, Germany}}
\affil[2]{\footnotesize{Deutsches GeoForschungsZentrum GFZ, Potsdam, Germany}}
\affil[3]{\footnotesize{Institute of Geosciences, Energy, Water and Environment, Polytechnic University of Tirana, Albania }}
\affil[4]{\footnotesize{Institute for Geological Sciences, Freie Universität Berlin, Germany}}
\affil[*]{\footnotesize{corresponding author}}

\date{Submitted: May 23, 2022}

\begin{document}
\renewcommand{\abstractname}{Summary}

\maketitle


\begin{abstract}
Machine Learning (ML) methods have demonstrated exceptional performance in recent years when applied to the task of seismic event detection. 
With numerous ML techniques now available for detecting seismicity, applying these methods in practice can help further highlight their advantages over more traditional approaches. 
Constructing such workflows also enables benchmarking comparisons of the latest algorithms on practical data. 
We combine the latest methods in seismic event detection to analyse an 18-day period of aftershock seismicity for the $M_{w}$ 6.4 2019 Durrës earthquake in Albania. 
We test two phase association-based event detection workflows methods, 
the EarthQuake Transformer (EQT; \citeauthor{mousavi2020earthquake}, \citeyear{mousavi2020earthquake}) end-to-end seismic detection workflow, and the PhaseNet \citep{zhu2019phasenet} picker with the Hyperbolic Event eXtractor \citep{woollam2020hex} associator. 
Both ML approaches are benchmarked against a data set compiled by two independently operating seismic experts who processed a small subset of events of this 18-day period. 

In total, PhaseNet \& HEX identifies 3,551 events, and EQT detects 1,110 events with the larger catalog (PhaseNet \& HEX) achieving a magnitude of completeness of $\sim$1. 
By relocating the derived catalogs with the same minimum 1D velocity model, we calculate statistics on the resulting hypocentral locations and phase picks. 

We find that the ML-methods yield results consistent with manual pickers, with bias that is no larger than that between different pickers. The achieved fit after relocation is comparable to that of the manual picks but the increased number of picks per event for the ML pickers, especially PhaseNet, yields smaller hypocentral errors.  
The number of associated events per hour increases for seismically quiet times of the day, and the smallest magnitude events are detected throughout these periods, which we interpret to be indicative of true event associations. 
\end{abstract}

\newpage

\section{Introduction}
\subsection{The seismic event detection problem: How is it tackled?}
Throughout recent years there has been an explosion in interest in the application of Machine Learning (ML) methods in seismology. 
Driven by increases in computational storage and compute capacity, these techniques are proving effective in solving a variety of tasks across the field. 
One such area which has received a particular interest is the task of seismic phase and event detection. 
Numerous approaches utilising ML are now attempting to detect seismic events at ever lower-signal to noise ratios. 
With this task forming a fundamental step to many seismological workflows, even minor increases in the event detection rate could greatly affect the subsequent understanding of any underlying physical-processes. 

As the detection of seismic events is an intrinsic point of many workflows in seismology, numerous methodologies have evolved for performing this task over decades. 
These varying approaches to detect the coherent energy of seismic events can be typically split into three categories: migration-based, cross-correlation-based, and phase association-based methods. 
The first two groups operate directly on time series, compromising either the (usually) filtered waveforms or characteristic functions, with the final group operating on pre-determined arriving phase information. 

Migration-based approaches detect events through coherency of the seismic wavefield, backpropagating waveforms or characteristic functions derived from them to find the focusing energy point in time and space as the source location and origin time \citep[e.g.][]{kao2004source, grigoli2013automated,drew13coalescence}. 
Whilst this approach enables the detection of smaller magnitude events due to the underlying stacking process, it is computationally expensive. 
Depending on the frequency range used and whether oscillatory waveforms, or unsigned characteristic functions are back-projected, this can also impose strict demands on the quality of the assumed velocity model. 
Additionally, if the seismicity rate increases, distinguishing between multiple events and their associated P- and S-wave radiated energy becomes challenging because many secondary maxima can occur due to interference of waves from different events, making it hard to understand which maxima correspond to real events. 

Cross-correlation-based event detection routines use the similarity of waveform signature to cluster waveforms belonging to the same seismic source region into groups. 
The predominant technique is template matching \citep{gibbons2006detection, shelly2007non} where a correlation function is used as the similarity metric, 
but more varied approaches using the same concept also exist \citep{perol2018convolutional, mousavi2019unsupervised, mcbrearty2019pairwise}. 
However, due to bias towards templates of existing seismicity, events at 'hot-spots' can be detected well but the analysis might be blind to earthquakes not located in already known clusters.

\begin{figure}
    \centering
    \includegraphics[width=.8\linewidth]{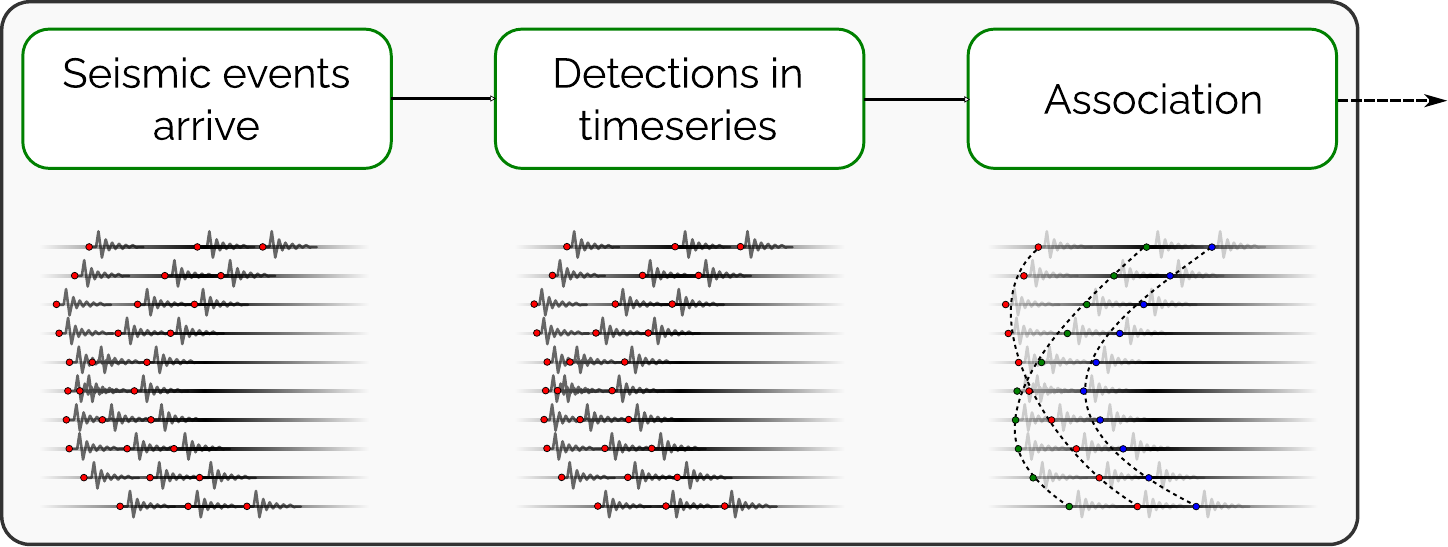}
    \caption{Schematic overview of a typical event detection pipeline.}
    \label{fig_ev_detect_schematic}
\end{figure}

Phase association-based event detection involves splitting the event detection pipeline into two distinct stages, highlighted in Figure \ref{fig_ev_detect_schematic}. 
Firstly, the impulsive onsets of seismic phase arrivals are detected (the seismic picking stage, Figure \ref{fig_ev_detect_schematic}) and classified via wave type (e.g. P- or S-wave). 
These independent detections of seismic energy are then correlated to their underlying source (the association stage, Figure \ref{fig_ev_detect_schematic}). 
The association step is non-trivial, as other sources of seismic energy arriving in the continuous data can result in false picks, complicating the association process. 

In this work we focus on the this specific approach, comparing the relative merits of phase association-based event detection workflows for analysing continuous data from a temporary aftershock deployment.

For many years, due to the heterogeneous nature of seismic wave propagation, the most accurate method was for a human expert to perform the picking stage. 
The natural drawbacks of having to employ a human expert is that the task becomes time-intensive, rendering the approach impractical in the current era of extensive data \citep{li2018high}.
In contrast, automated phase association pipelines offer orders of magnitude faster processing speeds, historically at the expense of event detection and location accuracy. 

In terms of methods, again there are a wide variety of techniques proposed. 
Automated picking algorithms can encompass traditional characteristic function-based approaches, applied for decades in real-time detection pipelines \citep{allen1978automatic70s, allen1982automatic80s, baer1987automatic, lomax2012automatic},  
but more recently, deep learning based picker have emerged as the leading automated picking method \citep{zhu2019phasenet, ross2018generalized, mousavi2020earthquake, Soto2021}. 
Deep learning routines are typically trained on millions of labelled phase examples to automatically infer the characteristic properties of seismic phase onsets. 
The latest deep learning routines now display accuracy levels similar to - or even exceeding - the performance of a human expert \citep{munchmeyer2022picker}. 
These methods are massively parallelized, and can be applied via  GPU architectures, enabling rapid processing. 
They are also able to detect significantly more picks compared to traditional approaches \citep{woollam2019convolutional}. 

Any increase in the phase detection rate renders the second component of an event association pipeline more difficult. 
Ignoring the potential for false picks, the simple power-law scaling of the Gutenberg-Richter relationship highlights that, as the minimum detectable event size decreases, orders of magnitude more events should be detected. 
Due to this factor, compute intensive phase association algorithms (e.g. backprojection-based, or cross correlation-based methods) will require significant compute time if they are to associate smaller magnitude events - which are key for enhancing the physical understanding \citep{ross2019searching}. 
This problem is further exacerbated by the increasing number of sensors present in the latest seismic deployments. 

To cope with increased level of information to correlate, there has also been a recent influx of novel association algorithms, which have been designed with scalability in mind. 
These can incorporate traditional ideas combined in a new way e.g. Rapid Earthquake Association and Location (REAL; \citeauthor{zhang2019rapid}, \citeyear{zhang2019rapid}), a hybrid approach which takes a typical waveform backprojection methodology and applies it in a 'sparse' way through using the coherency of detected phases instead of the waveform itself. 
There are also a more prominent general suite of techniques being applied to tackle the scalability problem; ML-based methods, which can utilise the information contained within extensive datasets to perform inference. 
Newly proposed phase association methods utilise a variety of ML techniques, 
from graph-theory \citep{mcbrearty2019earthquake}, 
Bayesian Gaussian Mixture Models for unsupervised clustering \citep{zhu2021earthquake},
recurrent neural networks \citep{ross2019phaselink}, and also
RANdom SAmple Consensus (RANSAC; \citeauthor{fischler1981random}, \citeyear{fischler1981random}), a data-driven ML technique to fit a parametric model to a data distribution  (e.g. \citeauthor{woollam2020hex}, \citeyear{woollam2020hex}; \citeauthor{zhu2021multi}, \citeyear{zhu2021multi}). 
These techniques aim to improve both accuracy and performance of the phase association task, decreasing the minimum threshold of detectable events given the new conditions of densely recorded picks in time and space. 

Accurate event association algorithms are therefore a key tool for seismologists aiming to better image and interpret the processes occurring within the subsurface.  
This work analyses 18 days of continuous data from a short period aftershock seismic network following the $M_{w}$ 6.4 Durrës event. 
We apply a selection of the latest ML-based methods for both seismic picking, and phase association to test how they perform associating a physical aftershock sequence recorded over a dense local array of recording instruments. 
Our analysis tests the PhaseNet \citep{zhu2019phasenet} and EQTransformer (EQT; \citeauthor{mousavi2020earthquake}, \citeyear{mousavi2020earthquake}) pickers. 
The EQT implementation also comes packaged with its own associator. 
This associator, along with the Hyperbolic Event eXtractor (HEX; \citeauthor{woollam2020hex}, \citeyear{woollam2020hex}) approaches are integrated into the detection pipelines to associate events.  
All detection pipelines are benchmarked against the manual event picks of two seismic experts who worked independently to analyse randomly selected events of the aftershock sequence. 
Such a workflow is similar in scope to \cite{cianetti2021comparison}, where we seek to evaluate the potential improvements ML algoirithms hold over traditional methods for event detection.

\subsection{Seismic setting}

\begin{figure}
    \centering
    \includegraphics[width=0.6\linewidth]{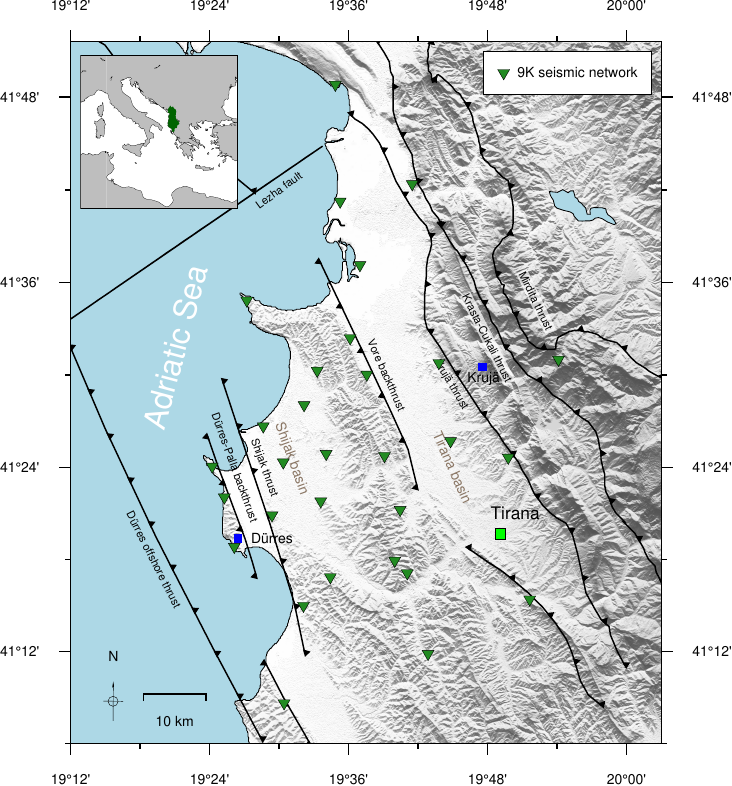}
    \caption{Seismic deployment following the $M_{w}$ 6.4 Durrës earthquake (2019-11-26). Temporary network shown with inverted triangles; stations are mix of 1 Hz and 4.5 Hz instruments. 
    The main faulting structures are  highlighted, geo-referenced from \citep{teloni2020seismogenic, vittori2020geological}. The Albania region is higlighted in the inset map in green.}.
    \label{fig:overview-mapplot}
\end{figure}

The continental collision of the Adriatic micro-plate with Eurasia generates the compressional tectonic structures observed throughout western Albania. 
Faulting structures in the region have been highly active, with notable events recorded in historical records. 
The largest of these being the 1979 $M_{w}$ 6.9 event occurring $\sim$100km north of the coastal city of Durrës (\citeauthor{benetatos2006finite}, \citeyear{benetatos2006finite}).
In recent years, the occurrence of numerous moderate-large magnitude events (2016, 2017, 2018) indicated a reactivation and changing stress dynamics within these faulting structures. 
The $M_{w}$ 6.4 mainshock ruptured $\sim$10 km north Durrës at a depth of 24 km \cite{gfz-261119-alb-eq-report} on 26 November 2019. 

The Geoforschungs Zentrum Postadam (GFZ) and Karlsruhe Institute of Technology (KIT) in conjunction with Polytechnic University of Tirana deployed 30 3-component short-period seismic instruments (a mixture of 1Hz and 4.5 HZ sensors) to record the aftershocks sequence (Figure \ref{fig:overview-mapplot}; \citeauthor{FDSN_9K_2019}, \citeyear{FDSN_9K_2019}). 
Stations were deployed from approximately two weeks after the mainshock. 
Stations were deployed close to houses to facilitate rapid deployment and ensure security of the recording devices which also means that an increased noise level had to be accepted. 
All stations were operated offline and the data was collected in regular visits of about 3 months, which became more complicated as the COVID-19 pandemic took hold in Europe.

\section{Methodology}
\subsection{Event detection methods}
To benchmark our various approaches, all methods are applied on the same period of continuous waveform data, 2019-12-13 to 2019-12-31. 
This covers an $\sim$18-day period directly following the deployment of the first instruments. 
With the first instruments from the temporary network only recording from 16 days after the mainshock, the recorded data does not capture the aftershock seismicity occurring immediately after the main rupture.  

\begin{figure}
    \centering
    \includegraphics[width=0.8\linewidth]{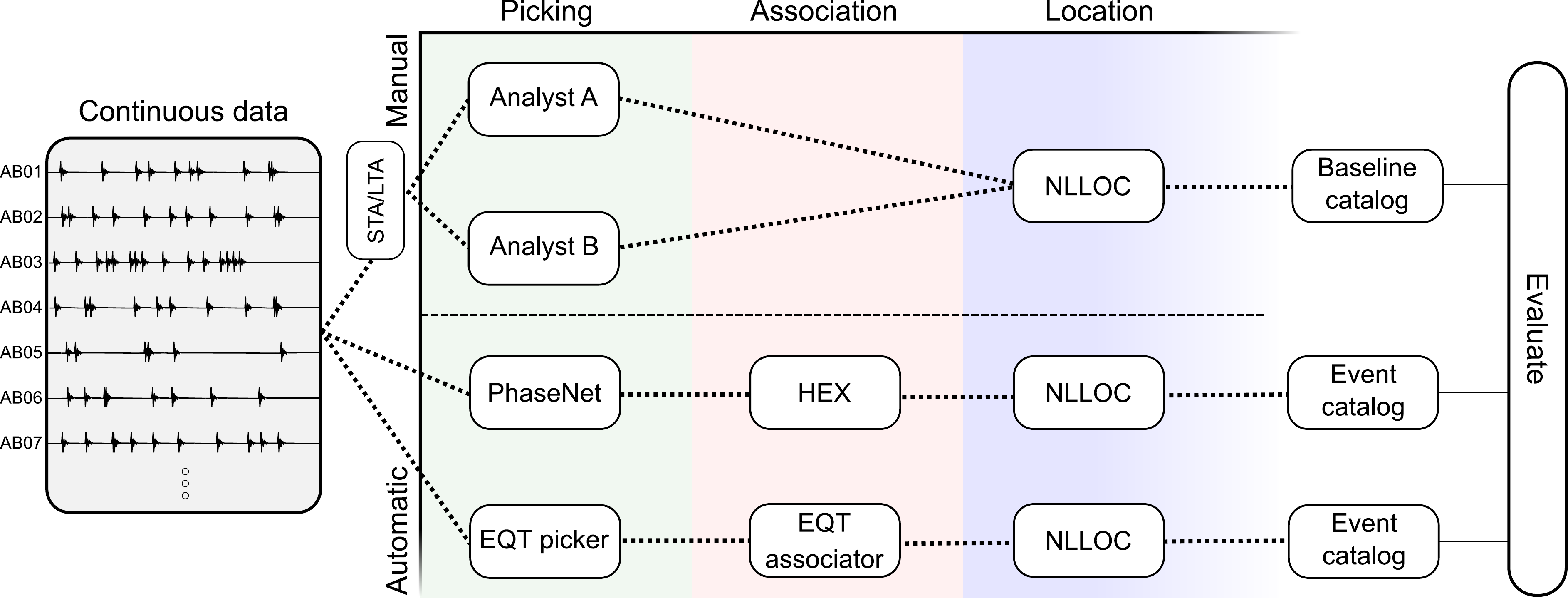}
    \caption{Outline of workflows for generating seismicity catalogues for comparing automated and manual approaches. All derived catalogs are evaluated against one another. See text for references to the different methods.}
    \label{fig:compare-workflow}
\end{figure}

The comparison of workflows for processing the 18 days of continuous waveform data are displayed in Figure \ref{fig:compare-workflow}. 
In total, there are three stages to our event association workflow: picking, association, and location. 
We apply three workflows, one combined manual approach, and two automatic approaches, resulting in three independent event catalogs over the analysis period. 
For the manually analysed control data set we applied a standard STA/LTA (short time average over long term average) arrival and event detection procedure as outlined in \cite{rietbrock2012aftershock}. 
The main reason behind this was to quickly compile a basic event list from which about 300 events were selected for manual analysis in order to gain a quick impression of aftershock activity and maybe identify the main causative fault.
The magnitude range of selected events is roughly between $M_{L}$ 1 and 4, providing a good range of magnitudes for the benchmarking exercise. 
For the automated procedures, our choice of algorithm for the association step was motivated by the results from the picking stage. 
Below, we first state the overall performance of the different picking methods. 

The total number of picks for each method are displayed in Table \ref{tab:pickstats}.
In total, analyst A picked 9,761 total phases, and analyst B picked 2,980 phases; for the automated approaches, EQT \citep{mousavi2020earthquake} picked 78,778 phases, and PhaseNet \citep{zhu2019phasenet} picked 561,326 phases. 
The total number of picks between manual approaches and automated approaches cannot be compared directly, however, as the manual picks these were not picked in a comprehensive manner.
Both EQT, and PhaseNet output probabilistic Characteristic Functions (CFs), which are a proxy for the probability of the occurrence of each phase. 
Picks are obtained from the output CFs by defining a cut-off value past which the sample of the maximum will be set as a pick onset. 
This threshold is a user-definable input parameter, for both picking algorithms we use the default value from the original publications.

\begin{table}
\captionof{table}{Picking statistics following the application of the various manual and automated approaches in processing all continuous waveform data over the recording network from 2019-12-13 to 2019-12-31. Values denote the total number of picks made in the first stage of the event detection pipeline, before any association stage.} \label{tab:pickstats}
\small
\centering
\begin{tabular}{ccccc}
\toprule
    Parameter & \multicolumn{2}{c}{Manual} & \multicolumn{2}{c}{Automatic} \\
    \cmidrule(lr){2-3}                  
    \cmidrule(lr){4-5}
     & Analyst A & Analyst B & EQT & PhaseNet \\
\midrule
P-picks & 5,758 & 1,680 & 39,918 & 289,509 \\
S-picks & 4,003 & 1,300 & 38,860 & 271,817 \\
total & 9,761 & 2,980 & 78,778 & 561,326 \\
\#P/\#S & 1.438 & 1.292 & 1.027 & 1.065 \\
\bottomrule
\end{tabular}
\end{table}

As the EQT pick catalog contains almost an order of magnitude less picks than the PhaseNet approach, irrespective of which cut-off probability threshold used, EQTs simple built-in association routine can deal with this level of pick information (one pick arriving on average every 20\,s across the network), whereas we use the machine learning-based HEX routine (\citeauthor{woollam2020hex}, \citeyear{woollam2020hex}) to associate the greater number of pick information generated by PhaseNet (one pick arriving on average every 3\,s, across the network) 
\footnote{Further statistics on relative performance of picking component as a function of varying cut-off thresholds are presented in Data Appendix, section A.1, to highlight how the EQT picker typically produced fewer picks than PhaseNet.}. 
In the final stage of each workflow the probabilistic location software NonLinLoc (NLLOC; \citeauthor{lomax2000probabilistic}, \citeyear{lomax2000probabilistic}) is applied to locate any associated events.

As the goal of this study is to verify the quality of automatic event catalogs made with the latest ML methods, any events located by each approach should, therefore, exhibit well-located hypocentres. 
\noindent To ensure all compared events are well-located, all final catalogs are filtered according to the following criteria. 
The minimum acceptable number of phases for a location are set as 6 for P-phases, and 4 for S-phases. 
As poor S-phase constraints are known to lead to incorrect focal depth estimates (e.g. \citeauthor{gomberg_effect_1990}, \citeyear{gomberg_effect_1990}), for the relocation step, we only relocate events with phase information satisfying the following criteria (\citeauthor{hardebeck_earthquake_2010}, \citeyear{hardebeck_earthquake_2010});  
azimuthal gap $<$ 180$^{\circ}$ (e.g. \citeauthor{kissling_geotomography_1988}, \citeyear{kissling_geotomography_1988}); 
at least 1 S-wave arrival must be at an epicentral distance less than the focal depth (e.g. \citeauthor{chatelain_microearthquake_1980}, \citeyear{chatelain_microearthquake_1980}).

\subsection{Evaluation}
We first compare the results of the two manual event catalogues against one another to see how they differ amongst themselves. 
This provides an assessment of how the baseline catalogs vary between one-another acts as a reference point of consistency for which the automatic procedures can be evaluated against. 
Following this comparison, the two manual event catalogues are merged to one combined manual event catalogue (without duplicates). 
Merging of matched events from manual analysts is done as follows: first, all picks and phases from both analysts are merged into a single combined manual catalog; 
then duplicated events are determined as events from each respective manual analyst with an onset time within 5s; 
finally, we fit an linear trend through the combined catalog events RMS relocation residual and total number of detected phases. 
As residual is strongly correlated to an event size and moveout across the network, factoring in this trend allows for determination of the 'better-quality' events when comparing matched events (discussed further in section \ref{ssec:matched-event-locs}). 
The event with the smallest 'corrected' RMS is then taken as the event to use. 
Only the phases of the 'best quality' manual events are then used in the combined catalog. 
This combined manual catalog is then compared to the EQT and PhaseNet catalogs (top workflow, Figure \ref{fig:compare-workflow}).

The quality of all automatic event locations are assessed by comparing against this suite of 'base' benchmark events.
Evaluating the latest events association methods against this approach provides a relative estimate against the 'traditional' state-of-the-art, when performing event association in practice. 
Once the varying manual and automated events catalogs are obtained, the workflow to pairwise compare events between catalogues is the following:

\begin{enumerate}
  \item  Find commonly detected events in both catalogues (using origin time). We consider events with an origin time difference of less than 2\,s to be the same event.
  \item Find common phase arrivals between pairs of matched events (using station and phase information) and calculate their residuals.
  \item Analyse the comparison statistics of matched events and phases.
\end{enumerate}

\noindent We determine events with an origin time difference of less than 2s to be the same event.

\section{Results}
\subsection{Benchmarking automated event detection workflows}
\begin{figure}
    \centering
    \includegraphics[width=1\linewidth]{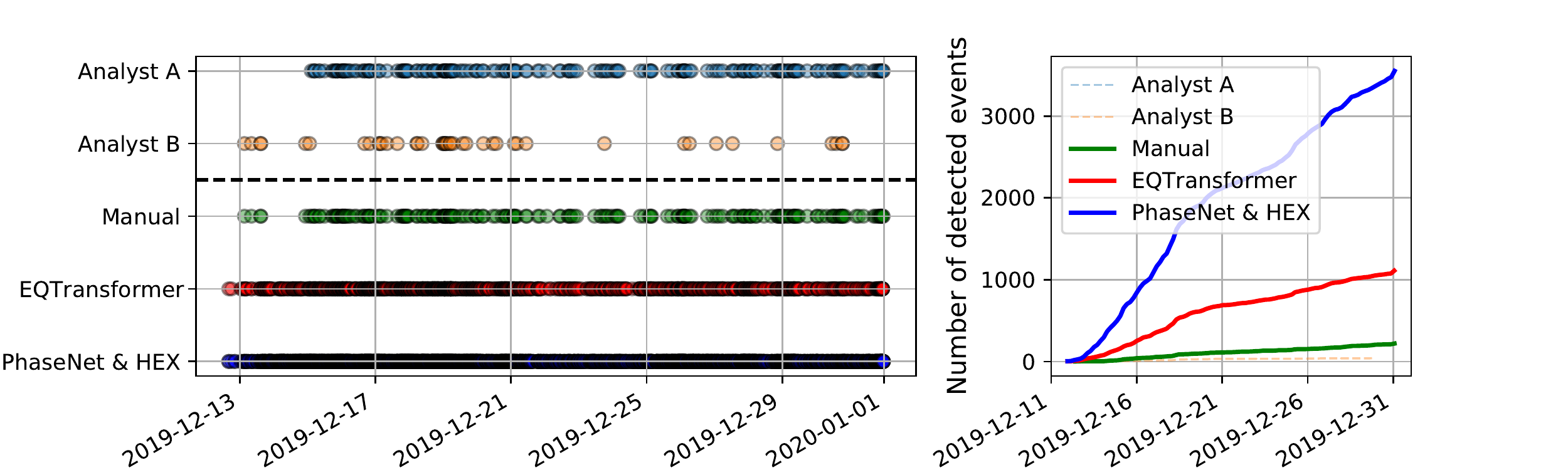}
    \caption{Event distribution in time. On the left each event is marked by a point, on the right the growth of cumulative number of events is shown.}
    \label{fig:tempextcats}
\end{figure}

The overall event detection results over the benchmarking period are displayed in Figure \ref{fig:tempextcats}, and Table \ref{tab:resulteventpickmatch}. 
From the combined manual baseline catalog, the EQT approach detects 94.1\% of the manual events, and the PhaseNet \& HEX approach detects 99.5 \% of the manual events. 
EQT finds 1,110 events in total, with PhaseNet \& HEX finding 3,551 events (Figure \ref{fig:tempextcats}).

\begin{table}
\captionof{table}{Comparison of the statistics of various automatic and manual event catalogs. Compared statistics are related to the matched events only and include: the number of matched events; the detected picks within matched events, and the mean ratio of the number of P picks and the number of S picks per matched event. For the manual analyst comparison, percentages are stated as fraction of picks of analyst A. For the automatic vs. combined manual catalog comparisons, the percentages are stated as a fraction of the combined manual baseline catalogs picks. EQT. denotes the EQTransformer event detection method; PhN. denotes the PhaseNet \& HEX detection method; An. A and An. B denote Analyst A and analyst B, respectively; Man. denotes the combined manual baseline event catalog.} \label{tab:resulteventpickmatch}
\small
\centering
\begin{tabular}{cccccccccc}
\toprule
  Comparison & \multicolumn{3}{c}{Manual}     
    & \multicolumn{3}{c}{EQTransformer} & \multicolumn{3}{c}{PhaseNet \& HEX}    \\
    \cmidrule(lr){2-4}                  
    \cmidrule(lr){5-7}
    \cmidrule(lr){8-10}
 Catalog &{An. A} & {An. B} & {match} & {Man.} & {EQT.} & {match} & {Man.} & {PhN.} & {match}  \\
\midrule
events	& 209 &	43 & 33 & 219 &	1,110 &	206 (94.06\%) & 219 & 3,351 & 218 (99.54\%)\\
picks &	1,318 & 1,640 &	1,278 (96.97\%) & 7,799 & 7,892 & 6,494 (83.27\%) & 8,221 & 11,107 & 7,937 (95.90\%)\\
\#P/\#S  &	1.41 & 1.18 &  & 1.36 & 1.05 &  & 1.35 & 1.09 & \\
\bottomrule
\end{tabular}\par
\bigskip
\end{table}

\subsection{Statistics of matched events and phases}
The results of the comparison routine are shown in Table \ref{tab:eventresiduals}, 
where we calculate statistics of the residual differences of matched events between catalogs, determined for a set of common event parameters. 
For example, the mean number of P-picks for the PhaseNet vs. manual baseline comparison  ($\mu = -4.8211$) means that on average, PhaseNet picks 4.8211 more P-picks when comparing matched events of the PhaseNet catalog with the manual baseline catalog 
\footnote{The full underlying distributions of each parameter in their respective catalogs are included within Fig. \ref{fig:tradeoffplotmatched} in the Data Appendix.}.  
We calculate statistics for the following parameters; depth, latitude, longitude, Root-Mean-Square location error (RMS), and the Number of Picks (NPS). 
We compare the events detected by both manual analysts against one another, to provide a rough reference point for the consistency of the manual seismic phase picking. 

\begin{table}
\captionof{table}{Comparison of statistics for common events between catalogs. For the manual events, the differences are calculated as the derived results from Analyst A compared against the results of Analyst B. Subscripts 'eqt', 'phn', 'base' refer to EQTransformer, PhaseNet \& HEX, and manual baseline, respectively.} \label{tab:eventresiduals}
\small
\centering
\begin{tabular}{cccccccccc}
\toprule
  Comparison & \multicolumn{2}{c}{Manual}     
    & \multicolumn{2}{c}{EQTransformer} & \multicolumn{2}{c}{PhaseNet \& HEX}    \\
    \cmidrule(lr){2-3}                  
    \cmidrule(lr){4-5}
    \cmidrule(lr){6-7}
 & {$\mu_{A - B}$} & {$\sigma_{A - B}$} 
 & {$\mu_{base - eqt}$} & {$\sigma_{base - eqt}$} 
 & {$\mu_{base - phn}$} & {$\sigma_{base - phn}$} \\
\midrule
Depth (km) & -0.2530 &   0.5764 &  0.2146 &   0.6235 &  0.0161 &   0.4339 \\
Latitude (km) & -0.0046 &   0.0031 & -0.0005 &   0.0030 & -0.0007 &   0.0028 \\
Longitude (km) & -0.0006 &   0.0056 & -0.0010 &   0.0045 & -0.0005 &   0.0034 \\
RMS (s) & -0.1148 &   0.0517 & -0.0033 &   0.0576 & -0.0243 &   0.0496 \\
number of picks & -10.9394 &   5.1109 & -0.2233 &  10.0491 &-13.0229 &   7.9616 \\
number of  P picks & -3.2424 &&  2.1456 && -4.8211 &  \\
number of  S picks & -6.5455 && -2.5971 && -8.4174 &  \\
\bottomrule
\end{tabular}
\end{table}

\begin{figure}
    \centering
    \includegraphics[width=\textwidth]{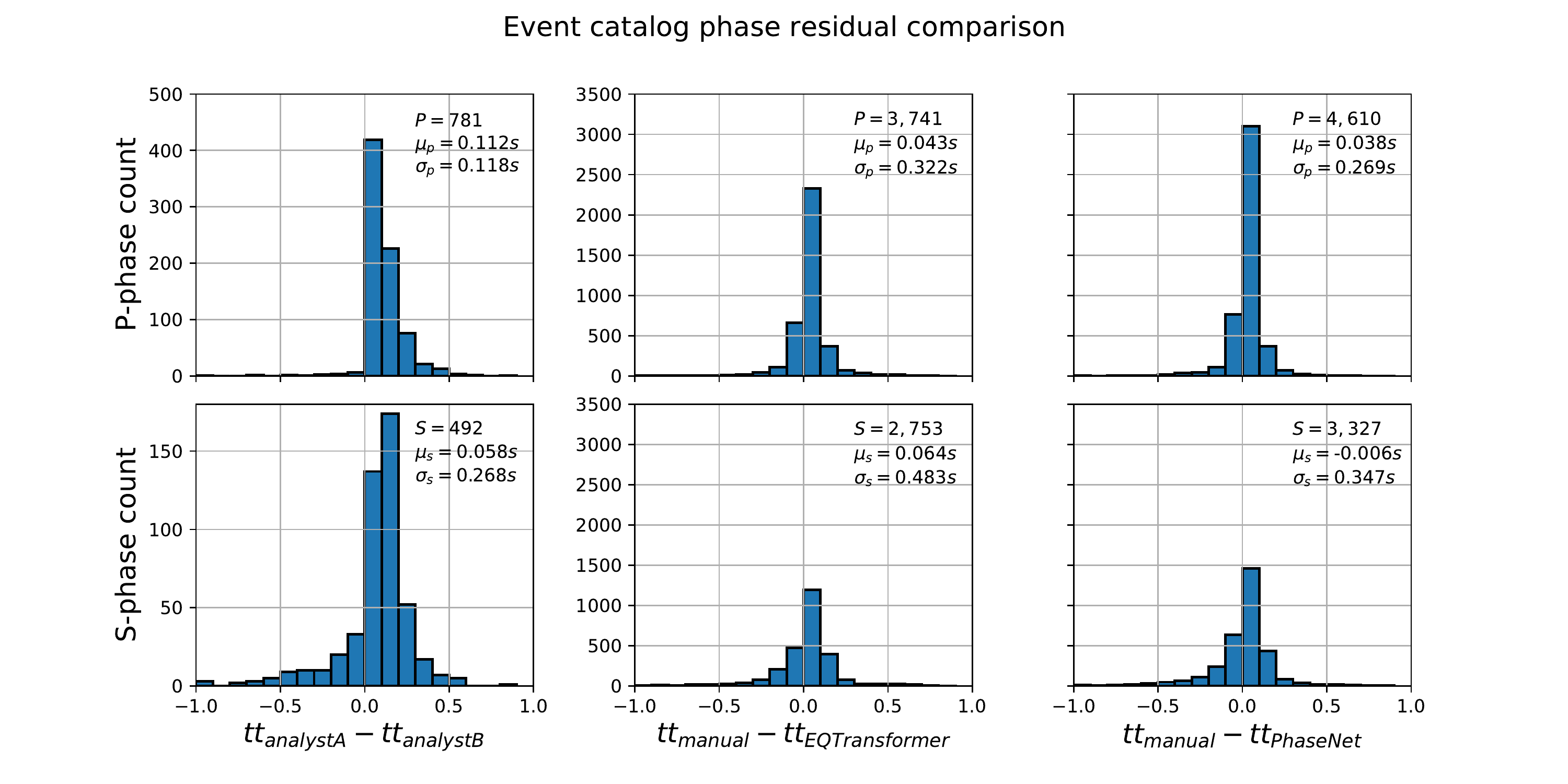}
    \caption{Phase arrival differences for matched phase picks of the matched events. The 'manual' comparison of Analyst A vs. Analyst B is the left-column, the EQTransformer vs. manual baseline comparison is the central column, and the PhaseNet \& HEX vs. manual baseline comparison is the right column.}
    \label{fig:phasettresiduals}
\end{figure}

For all matched events, the intersection of common picks are then examined.
Once common picks are identified between catalogs, the phase pick residuals can be determined, 
The statistics of pick differences are displayed in Table \ref{tab:pickresiduals} and Figure \ref{fig:phasettresiduals}. 
Residuals are calculated for arrival time in seconds, pick weight, and epicentral distance to the station where the phase was picked in kilometers. 
Again, all metrics describe the nature of the residual distribution when comparing matched phases in both catalogs. 
Traveltime denotes the difference in pick onset time. 
The underlying distributions for the traveltime residual comparison of matched phases are displayed in Figure \ref{fig:phasettresiduals}. 

The PhaseNet \& HEX, and EQT association approaches display an average difference in pick onset of 0.038, and 0.043 s, respectively, i.e., they pick slightly earlier than the baseline catalogue, which is dominated by the picks from analyst A. 
This average difference in pick onset between automated approaches and the baseline manual catalog is smaller than the mean difference between the two analysts, where analyst B seems to have picked  0.112 s earlier on average (this estimate is based on a very small number of events, though). 
However, the standard deviation of the pick time differences with respect to the manual baseline is larger for the automated methods (0.270 s for PhaseNet \& HEX, and 0.321 s for EQT) when compared against the variation observed between different analysts (0.122 s). 
So, whilst both automated approaches detect more picks per event (Table \ref{tab:eventresiduals}), these picks have  potentially larger associated uncertainties compared to a manual analyst.

\begin{table}
\small
\centering
\captionof{table}{Phase pick difference for matched phases. For the 'Manual' column the matched phase picks of Analyst A are compared with those of  Analyst B; for 'EQT' and 'PhaseNet' columns, the phase picks of EQTransformer and PhaseNet are compared against the manual baseline catalog, respectively. 'NLLoc weight' refers to the difference in pick weighting assigned in the NLLoc relocation procedure for matched picks.} \label{tab:pickresiduals}
\begin{tabular}{cccccccc}
\toprule
  Comparison & &\multicolumn{2}{c}{Manual }     
    & \multicolumn{2}{c}{EQTransformer} & \multicolumn{2}{c}{PhaseNet \& HEX}    \\
    \cmidrule(lr){3-4}                  
    \cmidrule(lr){5-6}
    \cmidrule(lr){7-8}
  & phase  & {$\mu_{A - B}$} & {$\sigma_{A - B}$} & {$\mu_{base - eqt}$} & {$\sigma_{base - eqt}$} & {$\mu_{base - phn}$} & {$\sigma_{base - phn}$}   \\
\midrule
Travel time (s) & P &    0.112 &    0.118 &    0.043 &    0.322 &    0.038 &    0.269 \\
& S &    0.058 &    0.268 &    0.064 &    0.483 &   -0.006 &    0.347 \\
NLLoc weight & P &    0.087 &    0.134 &   -0.066 &    0.156 &   -0.102 &    0.133 \\
& S &   -0.129 &    0.165 &   -0.002 &    0.196 &   -0.027 &    0.177 \\
\bottomrule
\end{tabular}
\end{table}

\subsection{Matched event locations}\label{ssec:matched-event-locs}
For evaluating the location consistency, the differences between the manual analysts can again provide a baseline metric which we can compare the automated catalogs against. 
The locations of compared event catalogs are displayed in Figure \ref{fig:compeventsonmap}.  
When comparing the 209 events of Analyst A and 43 events of Analyst B, 33 events are matched, detected across both event catalogues. 
These intersecting events differ in epicenter by 705 m on average (left panel, Figure \ref{fig:compeventsonmap}).
For the 206 matched events from the EQT catalog ($94.1\%$ of the total baseline events), 
these locations exhibit an average epicenter difference of 424 m when compared against the baseline catalog (centre panel, Figure \ref{fig:compeventsonmap}). 
Finally, 218 ($99.5\%$) of the base events are found using PhaseNet and HEX; 
locations show a difference of 377 m in epicentral distance on average (right panel, Figure \ref{fig:compeventsonmap}).

\begin{landscape}
\null
\vfill
\begin{figure}[!h]
    \centering
    \includegraphics[width=\linewidth]{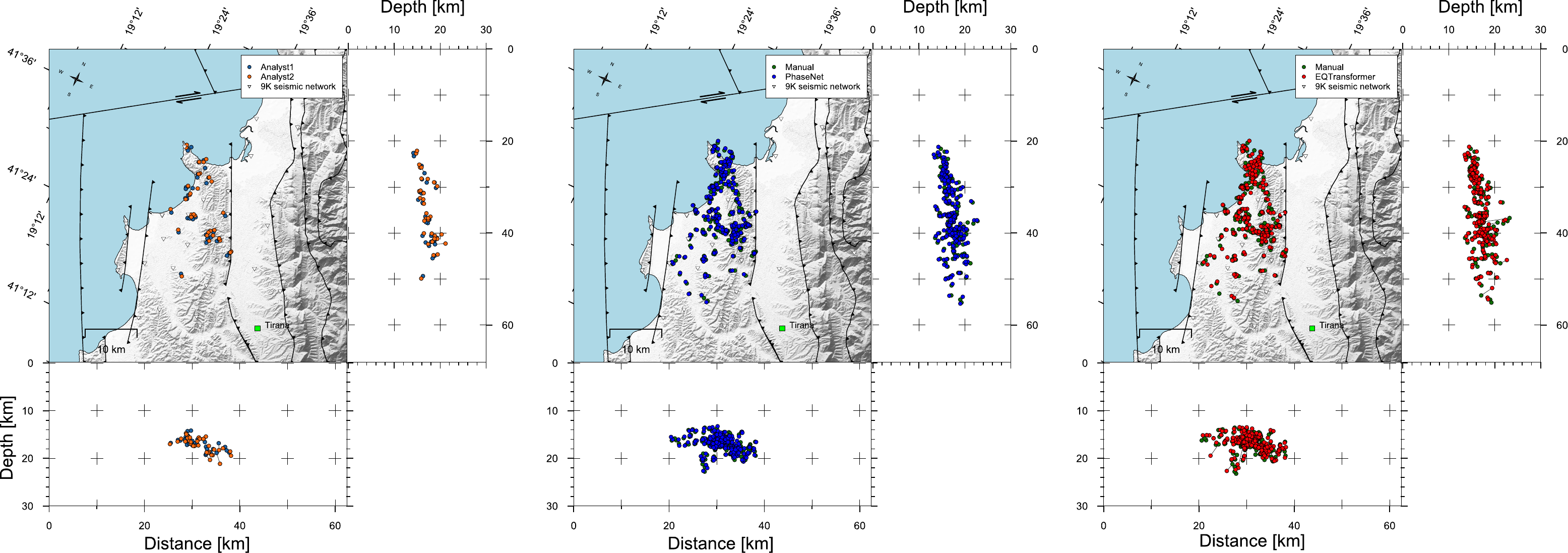}
    \caption{Matched event locations determined by Analyst A and Analyst B in the left panel; manual baseline and EQTransformer matched events in the centre panel, and manual baseline and PhaseNet \& HEX matched events in the right panel.}
    \label{fig:compeventsonmap}
\end{figure}
\vfill

\end{landscape}

\begin{figure}[h!]
    \centering
    \includegraphics[width=\textwidth]{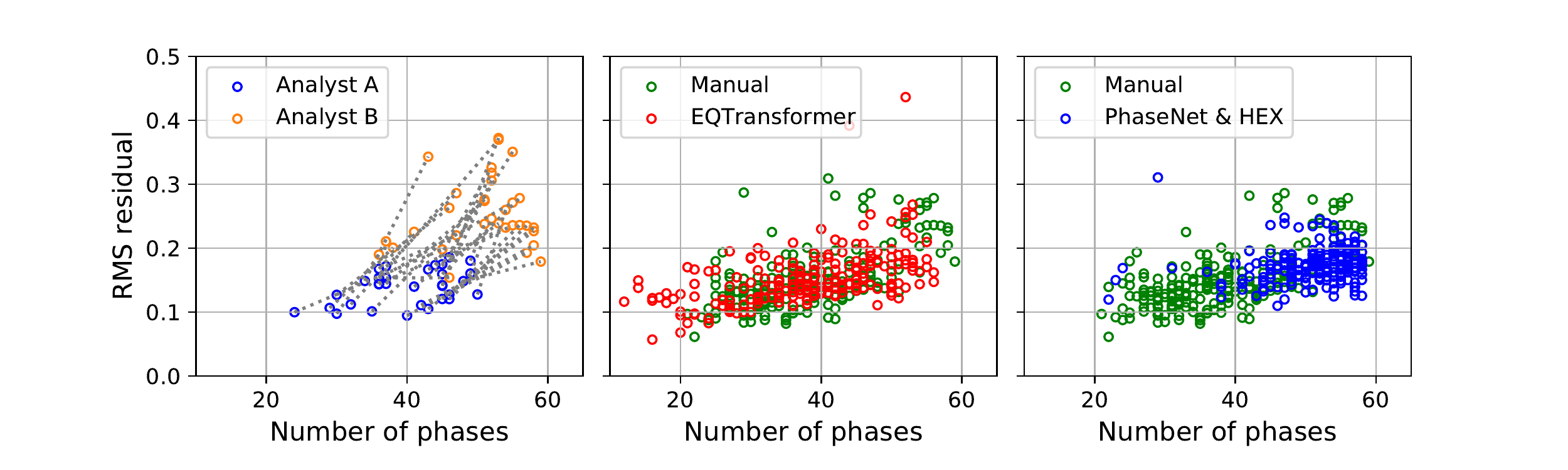}
    \caption{Travel time residual (RMS) as a function of number of picks for matched events from the manual Analyst A vs. Analyst B comparison (left), EQTransformer vs. Manual baseline catalog comparison (center), and PhaseNet \& HEX vs. manual baseline catalog comparison (right).}
    \label{fig:rmsnps}
\end{figure}

Figure \ref{fig:rmsnps} displays the travel  time  residual  (RMS) plotted  as  a  function  of  number  of  picks (NPS) for matched events in each of the event catalog comparisons.
Generally, we expect the RMS to increase with an increase in associated number of picks due to larger numbers of picks made at varying distance ranges incorporating higher pick uncertainties. 
This is easily visualised with the comparison between manual analysts. 
Analyst B picked more phases than Analyst A for every event, resulting in a larger RMS for every event, and is indicated by the positive slope of connecting line in left panel of Figure \ref{fig:rmsnps}. 
This signifies how we cannot simply take the standard RMS residual values as a 'gold standard' of uncertainty, as there are other contributing factors influencing the RMS. 
For example, not accounting for the 3D velocity perturbations in the relocation procedure could potentially account for the increase of RMS as a function of the number of picks; furthermore, picks at larger distances, more frequent for the automatic pickers, might suffer more from this. 

The automated EQT approach finds events with a similar RMS and number of associated picks to the manual baseline catalog (see central panel in Figure \ref{fig:rmsnps}). 
EQT makes $\sim$0.2 more picks per event on average (Table \ref{tab:eventresiduals}). 
In terms of location RMS, events located by the human analysts range between RMS 0.06 to 0.37 s, and for EQT event RMS ranges between 0.06 and 0.44 s (Figure \ref{fig:rmsnps}). 
Results for the Manual and PhaseNet \& HEX approach comparison are distinctively different. 
Here, we see that the majority of PhaseNet events have a RMS ranging between 0.11 and 0.31 s, and the number of picks per event ranges from 38 to 58. 
On average, the PhaseNet \& HEX method associate $\sim$13 more picks per event than the human analysts, increasing the RMS by 0.024 s on average (Table \ref{tab:eventresiduals}).

\section{Discussion}

The results of the benchmarking analysis indicate that the latest ML event detection methods at least match the reliability of individual analysts for individual events, while potentially being able to analyse very large number of events. 
More specifically, improvements do not only relate to greater numbers of event detections, but the ML detectors
can identify a greater number of phases per event, with the ability to pick at larger distance ranges with relatively minor increases in RMS residual. 
Both end-to-end automated workflows in the EQT and PhaseNet \& HEX approaches take continuous seismic data streams, and return $\sim$5x and $\sim$16x more events, respectively, than our manually analyzed catalog. 
219 events is a typical number of manually analyzed detections for temporary aftershock campaigns which can be quickly processed over a feasible time period. 
For a manual analyst to reach the number of detections of the ML workflows, this would be a significant, time-intensive undertaking.  
ML detected events are well-recorded across the seismic network, returning accurate event locations which can be used in subsequent analysis. 
Such results are in line with other works now applying ML tools to process seismicity \citep{tan2021machine, cianetti2021comparison}, where ML are displaying significant performance improvements over traditional methods. 
With the increasing popularity of ML event detection components, our results may also help inform future researchers who are seeking to  apply such techniques in and end-to-end fashion. 
Within the community, toolboxes are becoming available to streamline this task \citep{zhang2022locflow, woollam2022seisbench}, testament to their growing usefulness. 

Our investigation focuses on benchmarking a selection of fully integrated event detection pipelines, 
Crucially, the ML workflows presented here require no a-priori knowledge of the subsurface velocity structure to make these accurate detections. 
As the Albania region is one such area without a well-constrained velocity structure to date, 
we can see how applying such end-to-end detection workflows will greatly enhance understanding of the seismic structure of yet unexplored regions. 
This is especially true for the phase association problem, as whilst there might be room for performance increases by incorporating additional information like a region-specific 1D or 3D velocity model; however, the drawback of these approaches are that you are relying on a-priori knowledge, which might not be available for new regions or only of poor quality.
The full PhaseNet \& HEX event catalog from this work has been used in performing a new 1D velocity model inversion for the Albania region, 
using only results of the automated detection methods \citep{vanderheiden2021msc}. 

\begin{figure}[!h]
    \centering
    \includegraphics[width=.55\textwidth]{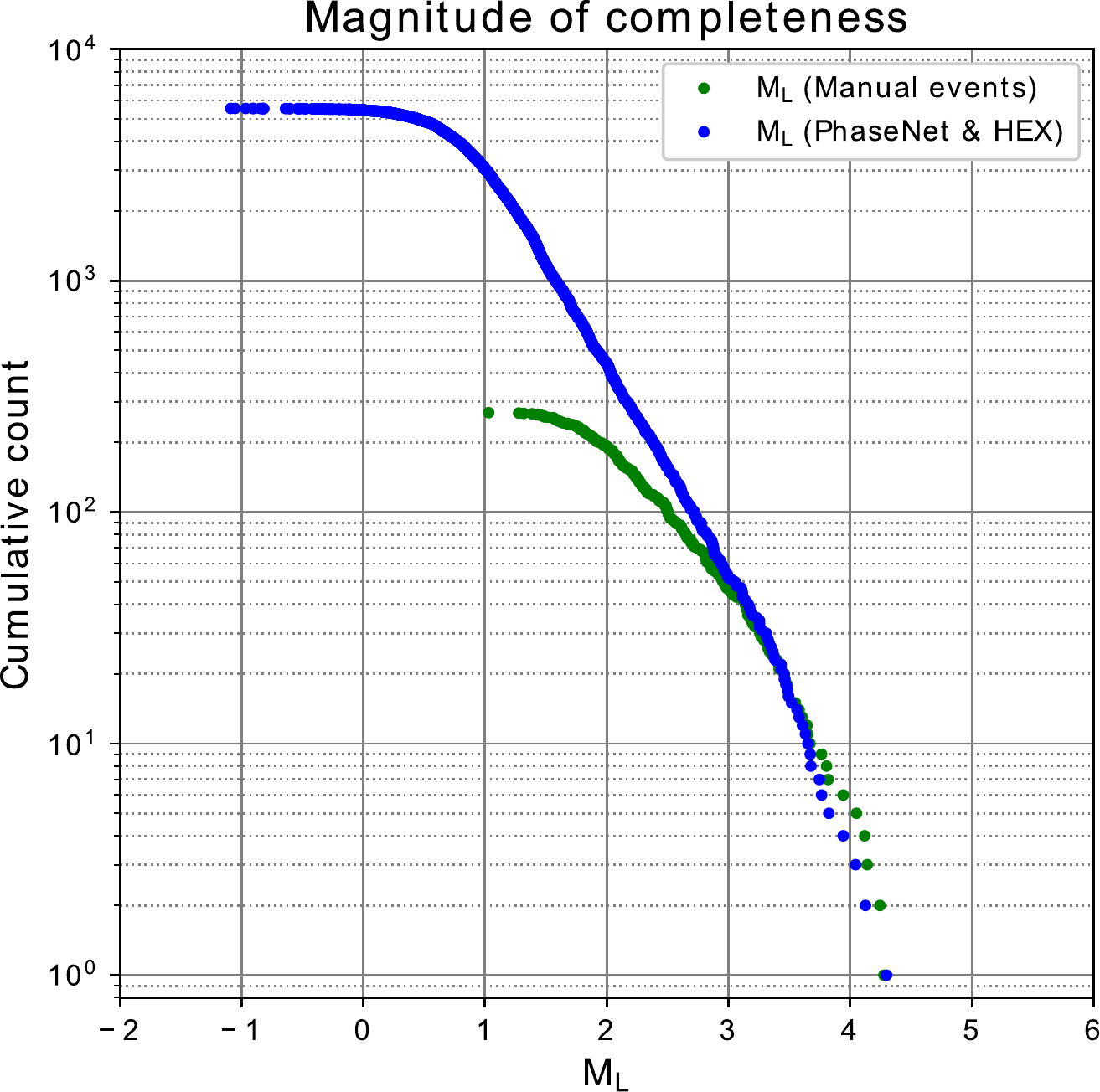}
    \caption{Frequency-magnitude distribution for the PhaseNet \& HEX detected events and manually detected events over the benchmarking period. The final quality location filtering has not been applied here.}
    \label{fig:mag-completeness-plot}
\end{figure}

One remaining question to ask is: Are all of the automatically detected aftershocks real events?
To answer this question we investigated in more detail the obtained catalogue by looking into the magnitude and the time distribution of the events. 
Firstly, we compute magnitude estimates ($M_L$) for all the events in both the manual, and the PhaseNet \& HEX catalog (Figure \ref{fig:mag-completeness-plot}), 
 using the calibration for the Albanian region of \cite{muco1991magnitude}.
We estimate magnitudes for all the events detected throughout the analysis period, so in this case, the final quality filtering step is not performed. 
This is as the magnitude calculation only very weakly depends on event depth, so here, we do not need such strict criteria, boosting the total number of associated events to 270 for the manual approach and 5,548 for the PhaseNet \& HEX approach. 
The PhaseNet \& HEX automatic catalog is determined to be complete above $M_{L}$ 1, with the manual catalog complete to $M_{L}$ 2.5. 
We note that the magnitude of completeness for the manual catalog should not be used in a direct comparison with the automated procedures, as the manual events were not picked in an exhaustive manner. 
It does, however, indicate what sort of completeness can be achieved when aiming to perform a standard manual analysis of an aftershock sequence, in a reasonable period of time. 

\begin{figure}[!h]
    \centering
    \includegraphics[width=.8\textwidth]{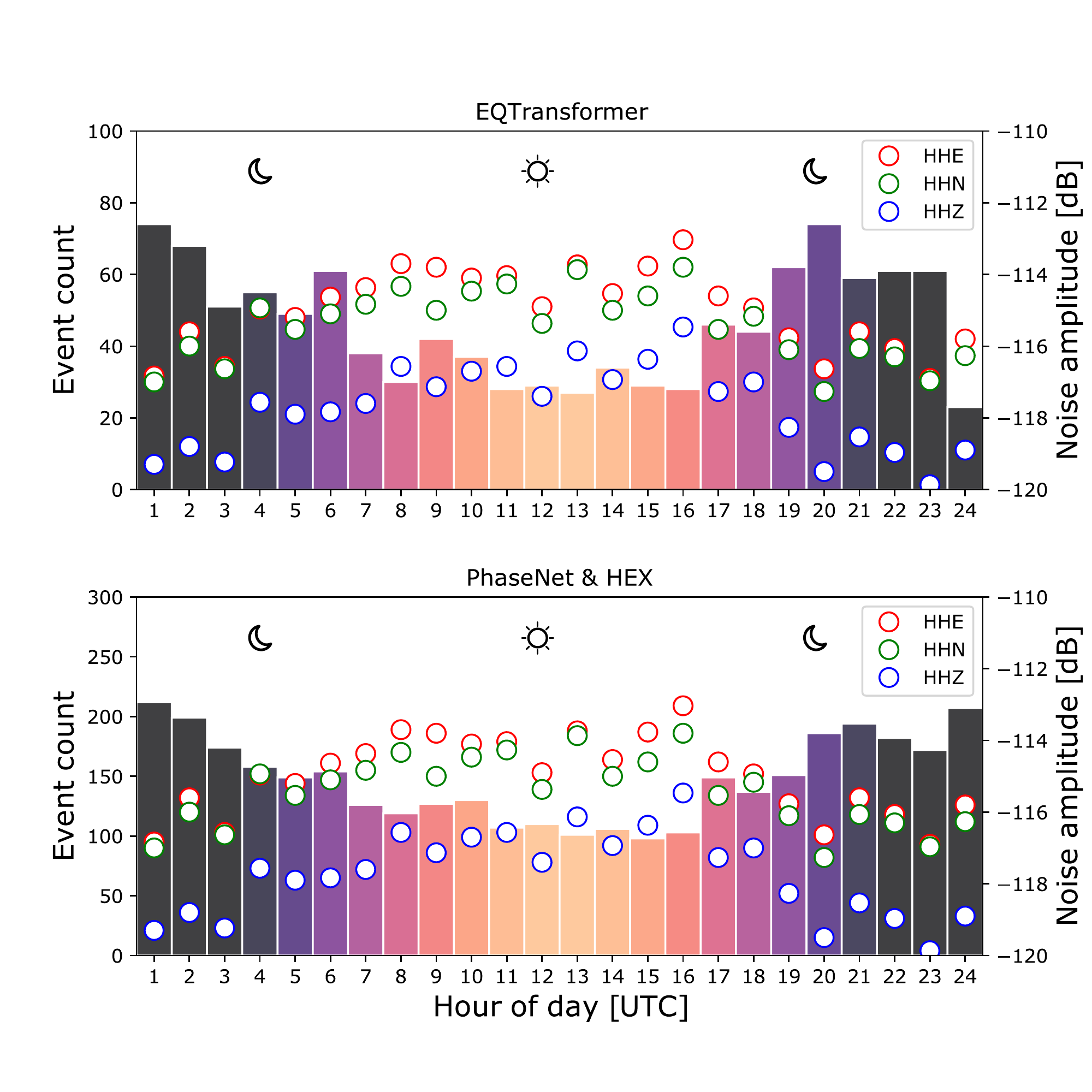}
    \caption{Number of detected aftershocks per hour in the day for both ML-based methods overlaid by the 3-component network noise in 1-10 Hz band.}
    \label{fig:eventsdiurnality}
\end{figure}

Secondly we investigate the dependence of the number of detected events on time of day in the benchmarking period. 
Figure \ref{fig:eventsdiurnality} displays the number of detected aftershock throughout the day for both ML-based methods. 
We also overlay the average amplitude of background noise (between 1 to 10 Hz) for each respective 3-component channel (HHE, HHN, HHZ).
A clear diurnal pattern is observed.
The noise estimation curve shows background noise increasing during the day and decreasing at night. 
As the seismic network is in the vicinity of populated areas, we can relate this to anthropogenic activity. 

\begin{figure}
    \centering
    \includegraphics[width=\textwidth]{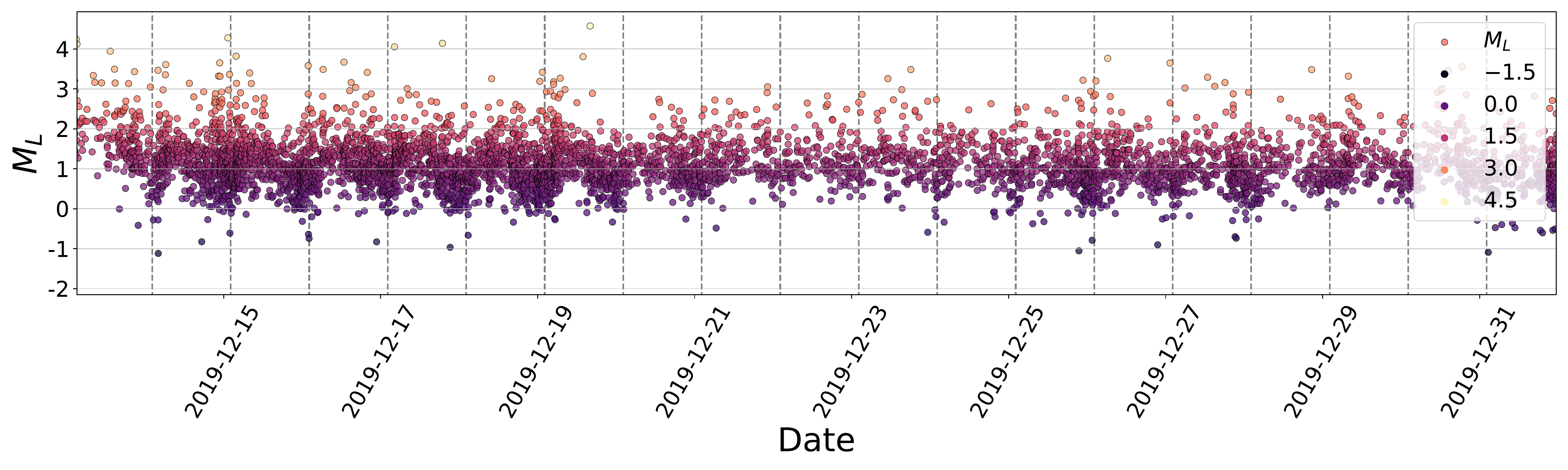}
    \caption{Magnitude-time plot of detected events over the benchmarking period for for the PhaseNet \& HEX detection workflow.}
    \label{fig:events-through-time}
\end{figure}

If the workflows were wrongly associating noise, we would expect the distribution of aftershocks throughout the day to match the daily background noise distribution, with more events in the middle of the day, during periods of higher background noise. 
For true associations, the expected behaviour would be to detect larger numbers of the smaller magnitude aftershocks during the night when the noise level drops (see the noise estimation point for components in the edge portions of Figure \ref{fig:eventsdiurnality}). 
The number of associated aftershocks through the day exhibits the latter behvaiour, indicating true associations (see the hourly event count bars in Figure \ref{fig:eventsdiurnality}). Such a time-of-day dependency of the number of detected events is well-known from manual analysis. 
This pattern is further highlighted when plotting the magnitude of events through time (Figure \ref{fig:events-through-time}). 
Here we can clearly see the local minima of detected event magnitudes roughly aligning with the occurrence of midnight.

Both methods are making true associations, uncorrelated to the background noise level but, a surprising result is the number of picks made by both ML methods in practice.
EQT makes 78,778 picks in total, and PhaseNet makes 561,326 total picks. 
This result is in contrast to the latest results of benchmarking works using curated benchmark datasets (e.g. \citeauthor{munchmeyer2022picker}, \citeyear{munchmeyer2022picker}), which determined EQT and PhaseNet to detect phases at a similar rate. 
One potential explanation is the length of the input window required for both algorithms. 
EQT requires a 60s input window, PhaseNet requires a 30s input window, but due to the nature of the CNN method along with a lack of global connections, its receptive field is $\sim$4\,s long. 
The larger input window for EQT means that longer term temporal dependencies can be optimised for during training. 
When deployed in practice to a new seismic region not seen during the training stage - such as Albania - these relatively longer-term temporal relationships may now differ due to the natural heterogeneity of wave propagation. 
Another way in which the learned temporal relationships could differ is due to the potentially higher seismicity rate of a temporary aftershock deployment. 
In such a network, simultaneously occurring events are more likely, overlapping in time to change the arriving waveform signature. 
These factors could be potentially contributing to the decrease in pick identifications when compared against PhaseNet, which applies a CNN with a smaller input window to detect local relationships only.

\section{Conclusions}
We have deployed two ML-based approaches to detect events occurring throughout an 18-day subset of the 2019 Durrës aftershock sequence. 
Both approaches are benchmarked against the results of manual analysts who selectively picked events during the same 18-day period of aftershock seismicity. 
From the two approaches, we find that the PhaseNet \& HEX method detects 3,551 events in total, and EQT detects 1,110 events in total, compared against the 219 manually analysed events.  

The difference in number of identified events between ML-methods is attributed to the step change in phase detections generated during phase picking. 
PhaseNet identifies 561,326 total picks, with EQT identifying 78,778 total picks when applied to detect continuous data in practice. 
By relocating the events using the same minimum 1D model, we matched events and phases common between seismic catalogs and analysed the resulting statistics. 
Further analysis indicates that the occurrence of the ML-detected events are inversely correlated to the background noise level, indicating true associations. 

\section*{Acknowledgements}
This work was funded as part of the REPORT-DL project under the grant agreement ZT-I-PF-5-53 and HEART. 
Instruments were provideded by the Geophysical Instrument Pool Potsdam, loan GIPP201928. Data are archived at the GEOFON data centre of the GFZ Potsdam, network code 9K (2019-2020) \citep{FDSN_9K_2019}.

\section*{Data availability statement}
Raw waveform data used in this study are archived at the GEOFON data centre of the GFZ Potsdam, network code 9K, restricted until 01/2024.

\newpage{}

\bibliographystyle{agsm}
\bibliography{bibliography}

@article{benetatos2006finite,
  title={Finite-fault slip models for the 15 April 1979 (Mw 7.1) Montenegro earthquake and its strongest aftershock of 24 May 1979 (Mw 6.2)},
  author={Benetatos, Christoforos and Kiratzi, Anastasia},
  journal={Tectonophysics},
  volume={421},
  number={1-2},
  pages={129--143},
  year={2006},
  publisher={Elsevier}
}

@MastersThesis{vanderheiden2021msc,
    author = {Van der Heiden, Vincent},
    title = {{Analysis of the 2019 $M_{w}$ 6.4 Albania aftershock sequence: An updated velocity model using AI-based solutions}},
    school = {Karlsruhe Insititute of Technology},
    address = {Geophysical Institute},
    year = {2021},
    }

@article{vittori2020geological,
    author = {Vittori, Eutizio and Blumetti, Anna Maria and Comerci, Valerio and Di Manna, Pio and Piccardi, Luigi and Gega, Dashamir and Hoxha, Ismail},
    title = "{Geological effects and tectonic environment of the 26 November 2019, Mw 6.4 Durres earthquake (Albania)}",
    journal = {Geophysical Journal International},
    volume = {225},
    number = {2},
    pages = {1174-1191},
    year = {2020},
    month = {12},
    issn = {0956-540X},
    doi = {10.1093/gji/ggaa582},
    url = {https://doi.org/10.1093/gji/ggaa582},
    eprint = {https://academic.oup.com/gji/article-pdf/225/2/1174/36567167/ggaa582.pdf},
}

@article{teloni2020seismogenic,
    author = {Teloni, Simone and Invernizzi, Chiara and Mazzoli, Stefano and Pierantoni, Pietro Paolo and Spina, Vincenzo},
    title = "{Seismogenic fault system of the Mw 6.4 November 2019 Albania earthquake: new insights into the structural architecture and active tectonic setting of the outer Albanides}",
    journal = {Journal of the Geological Society},
    volume = {178},
    number = {2},
    year = {2021},
    month = {02},
    issn = {0016-7649},
    doi = {10.1144/jgs2020-193},
    url = {https://doi.org/10.1144/jgs2020-193},
    note = {jgs2020-193},
    eprint = {https://pubs.geoscienceworld.org/jgs/article-pdf/doi/10.1144/jgs2020-193/5253113/jgs2020-193.pdf},
}

@article{zhang2022locflow,
    author = {Zhang, Miao and Liu, Min and Feng, Tian and Wang, Ruijia and Zhu, Weiqiang},
    title = "{LOC‐FLOW: An End‐to‐End Machine Learning‐Based High‐Precision Earthquake Location Workflow}",
    journal = {Seismological Research Letters},
    year = {2022},
    month = {03},
    issn = {0895-0695},
    doi = {10.1785/0220220019},
    url = {https://doi.org/10.1785/0220220019},
    eprint = {https://pubs.geoscienceworld.org/ssa/srl/article-pdf/doi/10.1785/0220220019/5565734/srl-2022019.1.pdf},
}

@article{woollam2022seisbench,
    author = {Woollam, Jack and Münchmeyer, Jannes and Tilmann, Frederik and Rietbrock, Andreas and Lange, Dietrich and Bornstein, Thomas and Diehl, Tobias and Giunchi, Carlo and Haslinger, Florian and Jozinović, Dario and Michelini, Alberto and Saul, Joachim and Soto, Hugo},
    title = "{SeisBench—A Toolbox for Machine Learning in Seismology}",
    journal = {Seismological Research Letters},
    year = {2022},
    month = {03},
    issn = {0895-0695},
    doi = {10.1785/0220210324},
    url = {https://doi.org/10.1785/0220210324},
    eprint = {https://pubs.geoscienceworld.org/ssa/srl/article-pdf/doi/10.1785/0220210324/5567643/srl-2021324.1.pdf},
}

@misc{gfz-261119-alb-eq-report,
  title="GFZ earthquake bulletin: Mw 6.4 Albania (26-11-2019).",
  author="{GFZ German Research Centre for Geosciences}",
  howpublished="\url{http://geofon.gfz-potsdam.de/eqinfo/event.php?id=gfz2019xdig}",
  year=2019,
  note="Accessed: 2022-03-28",
}

@article{FDSN_9K_2019,
  title={{AlbACa {\textendash} Albanian Earthquake Aftershock Campaign}},
  url={https://geofon.gfz-potsdam.de/doi/network/9K/2019},
  DOI={10.14470/4X7564679396},
  publisher={GFZ Data Services},
  author={Schurr, Bernd and Dushi, Edmond and Rietbrock, Andreas and Duni, Llambro},
  year={2020}
}

@article{cianetti2021comparison,
 title={{Comparison of Deep Learning Techniques for the Investigation of a Seismic Sequence: An Application to the 2019, Mw 4.5 Mugello (Italy) Earthquake}},
 volume={126},
 ISSN={2169-9356},
 url={http://dx.doi.org/10.1029/2021JB023405},
 DOI={10.1029/2021jb023405},
 number={12},
 journal={Journal of Geophysical Research: Solid Earth},
 publisher={American Geophysical Union (AGU)},
 author={Cianetti, S. and Bruni, R. and Gaviano, S. and Keir, D. and Piccinini, D. and Saccorotti, G. and Giunchi, C.},
 year={2021},
 month={Dec} }

@Article{drew13coalescence,
  Title                    = {Coalescence microseismic mapping},
  Author                   = {Drew, J. and White, R. S. and Tilmann, F. and Tarasewicz, J.},
  Journal                  = GJI,
  Year                     = {2013},
  Pages                    = {1773-1785},
  Volume                   = {195},

  Doi                      = {10.1093/gji/ggt331},
}

@article{rietbrock2012aftershock,
  title={Aftershock seismicity of the 2010 Maule Mw= 8.8, Chile, earthquake: Correlation between co-seismic slip models and aftershock distribution?},
  author={Rietbrock, A and Ryder, I and Hayes, G and Haberland, Christian and Comte, D and Roecker, S and Lyon-Caen, H},
  journal={Geophysical Research Letters},
  volume={39},
  number={8},
  year={2012},
  publisher={Wiley Online Library}
}

@article{kao2004source,
  title={The source-scanning algorithm: Mapping the distribution of seismic sources in time and space},
  author={Kao, Honn and Shan, Shao-Ju},
  journal={Geophysical Journal International},
  volume={157},
  number={2},
  pages={589--594},
  year={2004},
  publisher={Blackwell Publishing Ltd Oxford, UK}
}

@article{grigoli2013automated,
  title={Automated seismic event location by travel-time stacking: An application to mining induced seismicity},
  author={Grigoli, Francesco and Cesca, Simone and Vassallo, Maurizio and Dahm, Torsten},
  journal={Seismological Research Letters},
  volume={84},
  number={4},
  pages={666--677},
  year={2013},
  publisher={Seismological Society of America}
}

@article{Soto2021,
author = {Soto, Hugo and Schurr, Bernd},
doi = {10.1093/gji/ggab266},
issn = {0956-540X},
journal = {Geophysical Journal International},
month = {jul},
number = {2},
pages = {1268--1294},
publisher = {Oxford University Press},
title = {{DeepPhasePick: A method for detecting and picking seismic phases from local earthquakes based on highly optimized convolutional and recurrent deep neural networks}},
url = {https://academic.oup.com/gji/advance-article/doi/10.1093/gji/ggab266/6324012},
volume = {227},
year = {2021}
}

@article{tan2021machine,
  title={Machine-learning-based high-resolution earthquake catalog reveals how complex fault structures were activated during the 2016--2017 Central Italy sequence},
  author={Tan, Yen Joe and Waldhauser, Felix and Ellsworth, William L and Zhang, Miao and Zhu, Weiqiang and Michele, Maddalena and Chiaraluce, Lauro and Beroza, Gregory C and Segou, Margarita},
  journal={The Seismic Record},
  volume={1},
  number={1},
  pages={11--19},
  year={2021},
  publisher={Seismological Society of America}
}

@article{zhang2019rapid,
  title={Rapid earthquake association and location},
  author={Zhang, Miao and Ellsworth, William L and Beroza, Gregory C},
  journal={Seismological Research Letters},
  volume={90},
  number={6},
  pages={2276--2284},
  year={2019},
  publisher={Seismological Society of America}
}

@article{mcbrearty2019pairwise,
  title={Pairwise association of seismic arrivals with convolutional neural networks},
  author={McBrearty, Ian W and Delorey, Andrew A and Johnson, Paul A},
  journal={Seismological Research Letters},
  volume={90},
  number={2A},
  pages={503--509},
  year={2019},
  publisher={GeoScienceWorld}
}

@article{muco1991magnitude,
  title={Magnitude determination of near earthquakes for the Albanian network},
  author={Muco, B and Minga, P},
  journal={Bollettino di Geofisica Teorica ed Applicata},
  volume={33},
  number={129},
  pages={17--24},
  year={1991}
}

@article{fischler1981random,
  title={Random sample consensus: a paradigm for model fitting with applications to image analysis and automated cartography},
  author={Fischler, Martin A and Bolles, Robert C},
  journal={Communications of the ACM},
  volume={24},
  number={6},
  pages={381--395},
  year={1981},
  publisher={ACM New York, NY, USA}
}

@article{hardebeck_earthquake_2010,
	title = {Earthquake location accuracy},
	url = {http://www.corssa.org/export/sites/corssa/.galleries/articles-pdf/Husen-Hardebeck-2010-CORSSA-Eqk-location.pdf},
	doi = {10.5078/CORSSA-55815573},
	language = {en},
	urldate = {2021-03-14},
	author = {Hardebeck, Jeanne and Husen, Stephan},
	year = {2010},
	note = {Publisher: Community Online Resource for Statistical Seismicity Analysis},
}

@article{kissling_geotomography_1988,
	title = {Geotomography with local earthquake data},
	volume = {26},
	number = {4},
	journal = {Reviews of Geophysics},
	author = {Kissling, Edi},
	year = {1988},
	note = {Publisher: Wiley Online Library},
	pages = {659--698},
}

@article{chatelain_microearthquake_1980,
	title = {Microearthquake seismicity and fault plane solutions in the {Hindu} {Kush} region and their tectonic implications},
	volume = {85},
	number = {B3},
	journal = {Journal of Geophysical Research: Solid Earth},
	author = {Chatelain, Jean-Luc and Roecker, S. W. and Hatzfeld, D. and Molnar, Pm},
	year = {1980},
	note = {Publisher: Wiley Online Library},
	pages = {1365--1387},
}

@article{gomberg_effect_1990,
	title = {The effect of {S}-wave arrival times on the accuracy of hypocenter estimation},
	volume = {80},
	number = {6A},
	journal = {Bulletin of the Seismological Society of America},
	author = {Gomberg, Joan S. and Shedlock, Kaye M. and Roecker, Steven W.},
	year = {1990},
	note = {Publisher: The Seismological Society of America},
	pages = {1605--1628},
}

@article{gibbons2006detection,
  title={The detection of low magnitude seismic events using array-based waveform correlation},
  author={Gibbons, Steven J and Ringdal, Frode},
  journal={Geophysical Journal International},
  volume={165},
  number={1},
  pages={149--166},
  year={2006},
  publisher={Blackwell Publishing Ltd Oxford, UK}
}

@article{ross2019phaselink,
  title={PhaseLink: A deep learning approach to seismic phase association},
  author={Ross, Zachary E and Yue, Yisong and Meier, Men-Andrin and Hauksson, Egill and Heaton, Thomas H},
  journal={Journal of Geophysical Research: Solid Earth},
  volume={124},
  number={1},
  pages={856--869},
  year={2019},
  publisher={Wiley Online Library}
}

@article{mcbrearty2019earthquake,
  title={Earthquake arrival association with backprojection and graph theory},
  author={McBrearty, Ian W and Gomberg, Joan and Delorey, Andrew A and Johnson, Paul A},
  journal={Bulletin of the Seismological Society of America},
  volume={109},
  number={6},
  pages={2510--2531},
  year={2019},
  publisher={Seismological Society of America}
}

@incollection{lomax2000probabilistic,
  title={Probabilistic earthquake location in 3D and layered models},
  author={Lomax, Anthony and Virieux, Jean and Volant, Philippe and Berge-Thierry, Catherine},
  booktitle={Advances in seismic event location},
  pages={101--134},
  year={2000},
  publisher={Springer}
}

@article{zhu2021multi,
  title={A multi-channel approach for automatic microseismic event association using ransac-based arrival time event clustering (ratec)},
  author={Zhu, Lijun and Chuang, Lindsay and McClellan, James H and Liu, Entao and Peng, Zhigang},
  journal={Earthquake Research Advances},
  pages={100008},
  year={2021},
  publisher={Elsevier}
}

@article{woollam2020hex,
  title={HEX: Hyperbolic event extractor, a seismic phase associator for highly active seismic regions},
  author={Woollam, Jack and Rietbrock, Andreas and Leitloff, Jens and Hinz, Stefan},
  journal={Seismological Society of America},
  volume={91},
  number={5},
  pages={2769--2778},
  year={2020}
}

@article{zhu2021earthquake,
  title={Earthquake Phase Association using a Bayesian Gaussian Mixture Model},
  author={Zhu, Weiqiang and McBrearty, Ian W and Mousavi, S Mostafa and Ellsworth, William L and Beroza, Gregory C},
  journal={arXiv preprint arXiv:2109.09008},
  year={2021}
}

@article{allen1978automatic70s,
  title={Automatic earthquake recognition and timing from single traces},
  author={Allen, Rex V},
  journal={Bulletin of the seismological society of America},
  volume={68},
  number={5},
  pages={1521--1532},
  year={1978},
  publisher={The Seismological Society of America}
}

@article{allen1982automatic80s,
  title={Automatic phase pickers: Their present use and future prospects},
  author={Allen, Rex},
  journal={Bulletin of the Seismological Society of America},
  volume={72},
  number={6B},
  pages={S225--S242},
  year={1982},
  publisher={The Seismological Society of America}
}

@article{lomax2012automatic,
  title={Automatic picker developments and optimization: FilterPicker—A robust, broadband picker for real-time seismic monitoring and earthquake early warning},
  author={Lomax, Anthony and Satriano, Claudio and Vassallo, Maurizio},
  journal={Seismological Research Letters},
  volume={83},
  number={3},
  pages={531--540},
  year={2012},
  publisher={Seismological Society of America}
}

@article{zhu2019phasenet,
  title={PhaseNet: a deep-neural-network-based seismic arrival-time picking method},
  author={Zhu, Weiqiang and Beroza, Gregory C},
  journal={Geophysical Journal International},
  volume={216},
  number={1},
  pages={261--273},
  year={2019},
  publisher={Oxford University Press}
}

@article{munchmeyer2022picker,
  title={Which picker fits my data? A quantitative evaluation of deep learning based seismic pickers},
  author={M{\"u}nchmeyer, Jannes and Woollam, Jack and Rietbrock, Andreas and Tilmann, Frederik and Lange, Dietrich and Bornstein, Thomas and Diehl, Tobias and Giunchi, Carlo and Haslinger, Florian and Jozinovi{\'c}, Dario and others},
  journal={Journal of Geophysical Research: Solid Earth},
  pages={e2021JB023499},
  year={2022},
  publisher={Wiley Online Library}
}

@article{woollam2019convolutional,
  title={Convolutional neural network for seismic phase classification, performance demonstration over a local seismic network},
  author={Woollam, Jack and Rietbrock, Andreas and Bueno, Angel and De Angelis, Silvio},
  journal={Seismological Research Letters},
  volume={90},
  number={2A},
  pages={491--502},
  year={2019},
  publisher={Seismological Society of America}
}

@article{ross2019searching,
  title={Searching for hidden earthquakes in Southern California},
  author={Ross, Zachary E and Trugman, Daniel T and Hauksson, Egill and Shearer, Peter M},
  journal={Science},
  volume={364},
  number={6442},
  pages={767--771},
  year={2019},
  publisher={American Association for the Advancement of Science}
}

@article{mousavi2019unsupervised,
  title={Unsupervised clustering of seismic signals using deep convolutional autoencoders},
  author={Mousavi, S Mostafa and Zhu, Weiqiang and Ellsworth, William and Beroza, Gregory},
  journal={IEEE Geoscience and Remote Sensing Letters},
  volume={16},
  number={11},
  pages={1693--1697},
  year={2019},
  publisher={IEEE}
}

@article{ross2018generalized,
  title={Generalized seismic phase detection with deep learning},
  author={Ross, Zachary E and Meier, Men-Andrin and Hauksson, Egill and Heaton, Thomas H},
  journal={Bulletin of the Seismological Society of America},
  volume={108},
  number={5A},
  pages={2894--2901},
  year={2018},
  publisher={Seismological Society of America}
}

@article{mousavi2020earthquake,
  title={Earthquake transformer—an attentive deep-learning model for simultaneous earthquake detection and phase picking},
  author={Mousavi, S Mostafa and Ellsworth, William L and Zhu, Weiqiang and Chuang, Lindsay Y and Beroza, Gregory C},
  journal={Nature communications},
  volume={11},
  number={1},
  pages={1--12},
  year={2020},
  publisher={Nature Publishing Group}
}

@article{baer1987automatic,
  title={An automatic phase picker for local and teleseismic events},
  author={Baer, M and Kradolfer, U},
  journal={Bulletin of the Seismological Society of America},
  volume={77},
  number={4},
  pages={1437--1445},
  year={1987},
  publisher={The Seismological Society of America}
}

@article{shelly2007non,
  title={Non-volcanic tremor and low-frequency earthquake swarms},
  author={Shelly, David R and Beroza, Gregory C and Ide, Satoshi},
  journal={Nature},
  volume={446},
  number={7133},
  pages={305--307},
  year={2007},
  publisher={Nature Publishing Group}
}

@article{perol2018convolutional,
  title={Convolutional neural network for earthquake detection and location},
  author={Perol, Thibaut and Gharbi, Micha{\"e}l and Denolle, Marine},
  journal={Science Advances},
  volume={4},
  number={2},
  pages={e1700578},
  year={2018},
  publisher={American Association for the Advancement of Science}
}

@article{li2018high,
  title={High-resolution seismic event detection using local similarity for Large-N arrays},
  author={Li, Zefeng and Peng, Zhigang and Hollis, Dan and Zhu, Lijun and McClellan, James},
  journal={Scientific reports},
  volume={8},
  number={1},
  pages={1--10},
  year={2018},
  publisher={Nature Publishing Group}
}

\newpage{}
\section*{Appendix}
\setcounter{figure}{0}
\renewcommand\thefigure{A.\arabic{figure}}

\subsection*{Section A.1}

\begin{figure}[!h]
    \centering
    \includegraphics[width=1\linewidth]{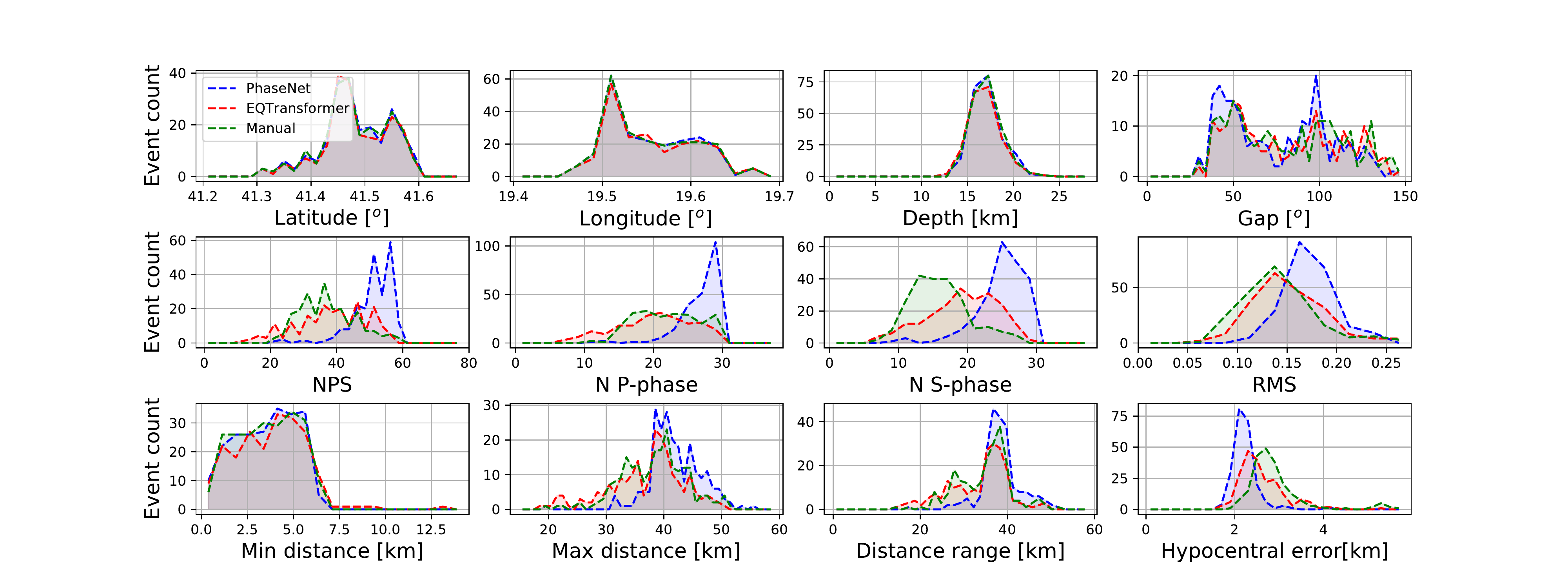}
    \caption{The matched events for each event catalog plotted as a function of varying parameters. The 'EQTransformer' label corresponds to the EQTransformer vs. manual baseline catalog event intersection, The 'PhaseNet \& HEX' label corresponds to the PhaseNet \& HEX vs. manual baseline catalog event intersection, the 'Manual' label corresponds to combined manual catalog Analyst A and Analyst B, containing 219 events. Hypocentral error is calculated as the length of the long axis of the confidence ellipsiod from location procedure output.}
    \label{fig:tradeoffplotmatched}
\end{figure}

\begin{figure}
    \centering
    \includegraphics[width=.6\linewidth]{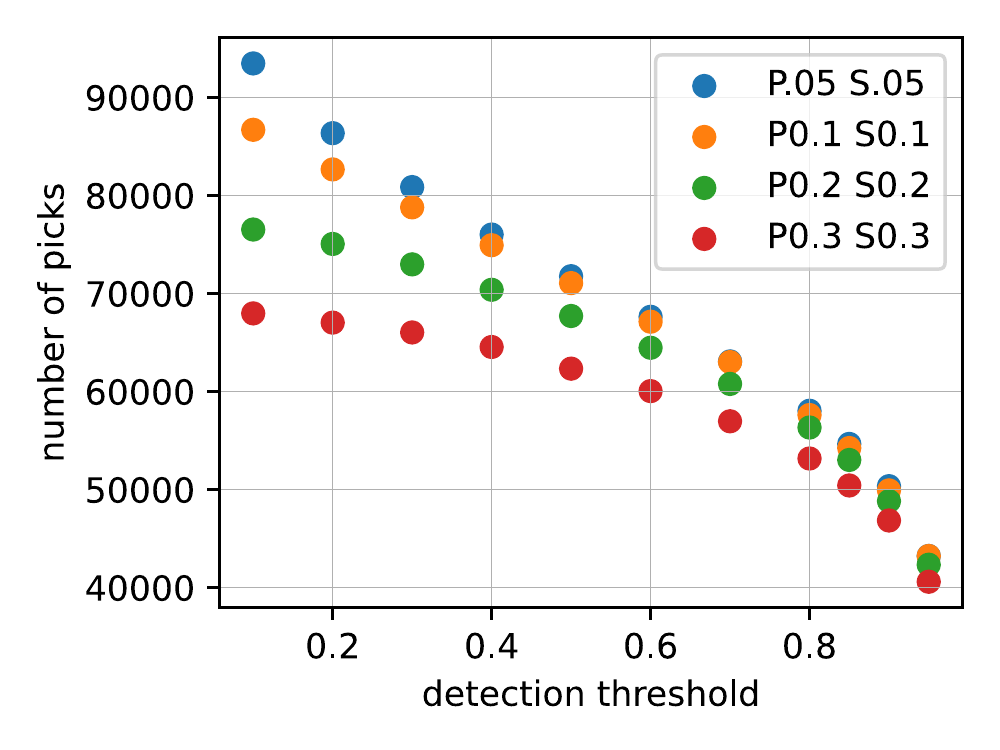}
    \caption{The total number of detected phases for EQTransformer, plotted as a function of the characteristic function/output probability cutoff detection threshold. The colorscale indicates the P-phase/S-phase probability cutoff threshold.}
    \label{fig:pick-detect-thresholds}
\end{figure}

\end{document}